\documentclass[preprint, amsmath,amssymb,aps, prb,superscriptaddress]{revtex4-1}
\usepackage{amsfonts,amssymb}
\usepackage{float}
\usepackage{graphicx}

\begin{document}
\draft
%

\title{Coulomb Gap in a Doped Semiconductor near the 
Metal-Insulator Transition: Tunneling Experiment and Scaling Ansatz}
\author{Mark Lee$\sp{(a)}$ and J. G. Massey}
\address{Department of Physics, University of Virginia, 
Charlottesville, VA 22903}
\author{V. L. Nguyen and B. I. Shklovskii}
\address {Theoretical Physics Institute, University of Minnesota,
Minneapolis, MN 55455}

\begin{abstract}
Electron tunneling experiments are used to probe Coulomb correlation 
effects in the single-particle density-of-states (DOS) of boron-doped 
silicon crystals near the critical density $n_c$ of the 
metal-insulator transition (MIT).  At low energies ($\varepsilon 
\leq$ 0.5 m$e$V), a DOS measurement distinguishes between insulating
and  metallic samples with densities 10 to 15 \% on either side of
$n_c$.  However, at higher energies ($\sim$ 1 m$e$V $\leq
\varepsilon \leq$ 50  m$e$V) the DOS of both insulators and metals
show a common behavior,  increasing with energy as $\epsilon^m$
where $m$ is roughly 0.5.  The observed characteristics of the DOS
can be understood using a  classical treatment of Coulomb
interactions combined with a  phenomenological scaling ansatz to
describe the length-scale dependence  of the dielectric constant as
the MIT is approached from the insulating  side.
\end{abstract}

\date{\today}
\maketitle
%
%
%

%
%
%
\section{Introduction}
\label{intro}
    Since 1979 the best available description of the disorder-driven 
metal-insulator transition (MIT) has been based on the 
non-interacting scaling theory of Abrahams,{\it et
al}.~\cite{Gang4}  Although it  treats thoroughly the localizing
effects of electron scattering  off static disorder, this theory is
incomplete because it  neglects electron-electron interactions.  In
particular, below  the critical density $n_c$ of the MIT the
vanishing carrier  mobility means that Coulomb correlations are
strong enough  to warrant treatment on equal footing with the
disorder.  The  importance of interactions is illustrated by several 
well-known, apparently anomalous phenomena,~\cite{Dobro} such as 
the recent discovery by Kravchenko, {\it et al.}~\cite{Krav95,Krav96} 
of a "forbidden" metallic state in a two-dimensional (2D) electron 
system.  In a disordered metal, electron-electron interactions lead
to  a singular negative correction to the single-particle 
density-of-states (DOS) near the Fermi level.~\cite{AA79}.  
Deep into the insulating side, it is well 
established that Coulomb correlations cause a 
Coulomb gap in the DOS near the Fermi level, which changes 
the temperature dependence of the DC electrical conductivity 
at very low temperatures.~\cite{ES75,ES84}  How these two 
renormalizations of the DOS evolve in the critical region of the
MIT and match each other at $n_c$ is one of the most challenging and 
long-standing questions of solid-state physics.  

    Combining both disorder and Coulomb interactions into a unified 
scaling description of disordered metals near the MIT was attempted 
by McMillan~\cite{MM81} and by Gefen and Imry~\cite{Gefen}.  These 
models were criticized~\cite{Lee82} for the use of the 
single-particle, rather than thermodynamic, DOS in describing the
charge  screening, as discussed in detail in Sect. III, and have not
been widely accepted despite garnering some experimental
support.~\cite{Hertel}   Later renormalization group theories by 
Finkelstein,~\cite{Fink83} Castellani, {\it et al}.,~\cite{Castel87} 
and Kirkpatrick and Belitz~\cite{KB90} used a Fermi-liquid 
based approach to describe the diffusion of interacting 
quasiparticles in a disordered metal.  These works employed 
a smooth thermodynamic DOS (or compressibility) and started from 
perturbative treatments of weak disorder.  All agree with the basic 
findings of Altshuler and Aronov~\cite{AA79} (AA) that there 
should be a square-root (in 3D) or logarithmic (in 2D) depletion 
in the single-particle DOS at the Fermi level of a disordered 
metal.  However, they failed to generate a well-defined, 
continuous charge localization transition at nonzero 
disorder and finite interaction strength, and cannot describe 
the emergence of a Coulomb glass state and the Coulomb gap as a 
system crosses over into the insulating state.  

    Tunneling experiments have observed AA-like depletions in the 
DOS of a variety of disordered metals, including amorphous metal-
semiconductor alloys,~\cite{McMo81,Ovadyahu,Hertel}, 
doped semiconductors,~\cite{Wolf71} and granular metals.~\cite{White}  
By contrast, with the exception of the singular sodium tungsten 
bronze system,~\cite{Davies} the existence of the Coulomb gap in
localized  insulators has until recently been only indirectly
inferred from  activation fits to DC conductivity~\cite{Rosenbaum} 
or relaxation measurements.~\cite{Monroe}  Only in the last 
few years have quantitative tunneling spectroscopic observations 
of the Coulomb gap been made in 3D~\cite{Massey95,Massey96} and 
in 2D~\cite{Ashoori} non-metallic semiconductors.  

    Most previous work on interaction effects has emphasized the 
metallic side; much less corresponding effort, experimental or 
theoretical, exists to describe the insulating side near the 
critical region of the MIT.  In this paper we concentrate on 
studying the Coulomb gap in the DOS that occurs in 3D disordered 
insulators close to the MIT.  Results of tunneling measurements 
of the DOS in boron-doped silicon are presented over a much 
larger range of energies than in Ref~\onlinecite{Massey95,Massey96}.  
The tunneling DOS spectra show that metals and insulators can be 
distinguished by the low-energy characteristics of the DOS 
({\it i.e.} square-root cusp vs. parabolic Coulomb gap), but that
metals and  insulators share a common, roughly square-root
high-energy DOS  behavior that is approximately independent of
dopant densities close  to $n_c$.  Of course, a truly complete
microscopic theory of the  MIT with interactions should cover the
continuous crossover from  insulating to metallic state.  We present
a much more modest and simple  scaling approach to the MIT, which
emphasizes the insulating side  and is based on an extension of the
Efros-Shklovskii~\cite{ES75}  arguments.  Using the idea of a
dielectric constant with  spatial dispersion,~\cite{MM81,Gefen} we
argue that  at the point of the MIT a power law depletion of the DOS
near  the Fermi level is still the ES Coulomb gap, only modified by 
this dispersion.  The scaling ansatz describes well the main 
features of the experimental data.  

    This paper is organized as follows: Section II presents 
experimental details and data, Section III develops the scaling 
description, and Section IV analyzes of the data with respect 
to the scaling model.

\section{Tunneling Measurements of the Coulomb Gap}
\label{Tunneling}
    When a conductor is separated from a conventional metal by a 
rectangular potential barrier high enough to prevent classical 
current flow but thin enough to permit quantum tunneling, the 
tunneling conductance at temperature $T$, $G(V,T) = dI/dV$ where 
$I$ is the tunneling current and $V$ is the voltage bias across 
the junction, is given by:~\cite{McR}
\begin{equation}
\frac{G(V)}{G_0}=\int \frac{N(\varepsilon)}{N_0}~\frac{\partial
f(\varepsilon - eV,T)}{\partial eV}~d\varepsilon
\label{tunnelconductance}
\end{equation}
where $G_0$ is the conductance in the non-interacting case, 
$N(\varepsilon ,T)/N_0$ is the interacting single-particle DOS 
relative to the non-interacting value $N_0$, and $f$ is the Fermi 
function.  We take the zero of energy at $\varepsilon_F = 0$.  The 
highest bias used in this experiment is 50 m$e$V, which is much 
smaller than the several $e$V height of the SiO$_2$ barrier, so that 
the barrier transmission coefficient is taken to be independent of 
bias.  In nearly all tunneling measurements, the normalizing
conductance $G_0$ is taken at a relatively high voltage bias, where
$G(V)$ is  either constant or only slowly varying on the energy
scale of the  spectral features of interest.  The normalized
conductance then  gives the ratio $N(eV,T)/N_0$, thermally broadened 
by convolution with $-\partial f/\partial (eV)$.  At sufficiently 
low temperature, when $k_BT$ is much smaller than the  energy scale 
of characteristic variations in $N(\varepsilon)$, ordinary thermal 
broadening can be neglected so that $G(V,T)/G_0$ is directly 
proportional to $N(eV,T)/N_0$.  Note that, unlike the non-interacting 
case, $N(eV,T)$ can have a non-trivial intrinsic temperature 
dependence separate from ordinary thermal broadening.

\begin{figure}[H]
\includegraphics[width=\linewidth]{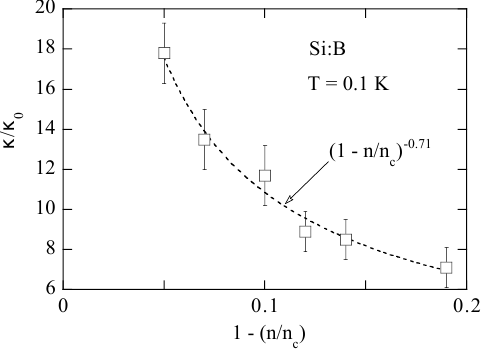}
\caption{Static dielectric constants (in units of the Si dielectric 
constant $\kappa_0 = 11.7$) as a function of the normalized dopant 
density for a set of insulating Si:B crystals.  These data were 
measured at a temperature $T$ = 0.1 K.  The solid curve is a 
fit to the functional form $\kappa (n/n_c) \sim 
(1 - n/n_c)^{-\zeta}$, which gives $\zeta \approx 0.71$.}
\label{kappa(n)}
\end{figure}

    The Si:B crystals used were characterized by measuring their 
room-temperature resistivities $\rho$ and resistivity ratios (RRs) 
$\rho$(4.2 K)/$\rho$(300 K).  Dopant densities $n$ were obtained 
using the calibration of Thurber, {\it et al.}~\cite{Thurber} and 
the data of Dai, {\it et al.}~\cite{Dai91} to translate between 
measured RRs and dopant densities.  Samples used had RRs of 
2.3 to 18, corresponding to $n/n_c$ of 110 \% to 81 \%, respectively, 
where we take~\cite{Dai91} $n_{c} = 4.0 \times 10^{18}$ cm$^{-3}$.  
Details of the DC variable-range hopping (VRH) conductivity in many 
of these samples have been published 
previously.~\cite{Massey95,Massey97}  The static dielectric
constants of several insulating crystals were  measured using a
capacitance method similar to that described  by
Castner.~\cite{Castner}  These results are shown in
Fig.~\ref{kappa(n)}.   To compare values with the literature, on an
86 \% sample we  obtained $\kappa /\kappa_0 = 8.5 \pm 1$, where
$\kappa_0 = 11.7$  is the dielectric constant of the host Si
lattice.  This value  is in reasonable agreement with published
values~\cite{Rosenbaum83}  on Si:P and Si:As at similar values of
$n/n_c$.   The curve in  Fig.~\ref{kappa(n)} is a fit of the data to
$\kappa (n) \sim (1 - n/n_c )^{-\zeta}$, which yields $\zeta \simeq
0.71$.

    Tunnel junctions were formed by creating a very thin (estimated 
15 to 30 \AA) layer of SiO$_2$ as a tunnel barrier on the Si:B 
crystals, and using an Al film as the counter-electrode.  Details 
of the fabrication of these metal-oxide-semiconductor structures were 
described previously.  The major modification from the prior 
description was the use of an ultraviolet ozone method in place of 
a chemical oxide growth to clean the silicon surfaces and 
produce more reproducible ultra-thin SiO$_2$ tunnel barriers.  
SiO$_2$ thicknesses were estimated from published calibrations of 
UV exposure time, temperature, and O$_2$ flow rate.~\cite{Fujitsu}  
Good junctions were those which showed a signature of the Al 
superconducting energy gap below 1 K.  These devices had junction 
resistances between 65 to 850 k$\Omega$ near 1 K.

    Tunneling current-voltage $(I-V)$ and conductance-voltage 
$(G(V))$ traces were taken using standard analog methods.  Where
required, a  small (1 kG) magnetic field was used to suppress the
superconductivity  in the Al electrode.  We shall refer to such data
as "zero Teslas."   Data down to 1.2 K were obtained by suspending
the samples stress-free  from their leads and immersing in a pumped
liquid $^4$He bath.  To  reach lower temperatures, the samples were
immersed in the 
$^3$He/$^4$He mixture of a dilution refrigerator.  Cooling power at 
0.1  K was measured to be 120 $\mu$W, while no more than 0.1 $\mu$W 
of power was used to take the data.  Temperature stability of better 
than 1 mK at 1.2 K and 0.1 mK at 0.1 K could be obtained, which was 
important to prevent thermal fluctuations from coupling to the highly 
temperature-dependent resistance in the more insulating samples.  For 
samples with $n/n_c >$ 90 \%, reliable data could be taken below 0.1 
K.  For insulating samples with $n/n_c <$ 90 \%, the resistance of 
the Si:B crystal itself rose to exceed 10 \% of the resistance
across the  tunnel junction below 0.1 K.  Because the crystal acted
as a voltage  drop in series with the junction, quantitative
tunneling  conductance data below $\sim$ 0.1 K is not reliable in
these most  insulating samples.  Therefore, we present only $T \geq$
0.5 K data  for these samples, where the voltage drop across the
Si:B crystal  is no more than a 1 \% correction to the tunnel
junction conductance.  

\begin{figure}[H]
\includegraphics[width=\linewidth]{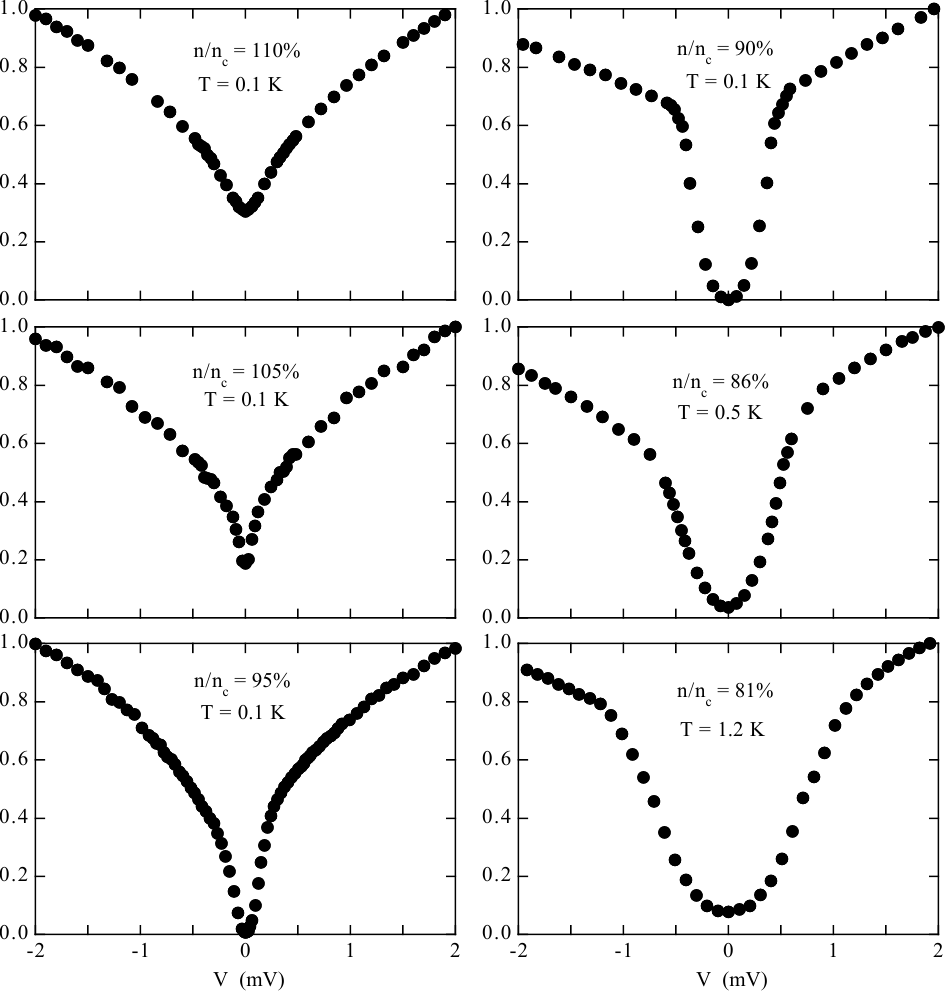}
\caption{Low-bias tunneling conductance-voltage spectra for Si:B 
samples at several different values of $n/n_c$.  All data sets are 
normalized to the conductance at +2 mV.  Data for samples with 
$n/n_c \geq$ 90 \% were taken at $T$ = 0.1 K.  Data for the 
86 \% and 81 \% samples were taken at 0.5 K and 1.2 K, respectively.}
\label{G(V)lowbias}
\end{figure}

    Figure~\ref{G(V)lowbias} shows low-energy tunneling conductance 
spectra on six samples with densities $n/n_c =$ 81 \%, 86 \%, 90 \%,
95 \%, 105 \%,  and 110 \%, all in zero Teslas.  All data are
normalized to the  conductance value of each sample at +2 mV.  The
spectrum for the  81 \% sample was taken at 1.2 K, the spectrum for
the 86 \% sample  was taken at 0.5 K, and the data for the other
four samples were  taken at 0.1 K.  Ordinary thermal broadening at
the measurement  temperatures is minor compared to the energy scale
of the  conductance features, and so has not been deconvoluted from 
the data.  However, the interacting DOS of each sample has an 
intrinsic finite temperature dependence that may be more obvious 
in the data for the 81 \% and 86 \% samples because of the higher 
temperature.  In the conductance spectra, there is a clear difference 
between the insulating and metallic samples at low biases, 
$|V| <$ 1 mV.  In all the insulating samples, a 
conductance dip across zero voltage (the Fermi level in tunneling 
measurements) is the signature of the Coulomb gap.  For the 81 \% 
and 86 \% samples, $G(V=0)$ is small but greater than zero because 
the measurement temperature for these two samples is sufficiently 
high that there is some intrinsic filling in of the $T = 0$ 
gap.  In all insulating samples, the low-energy conductance 
spectra fit a power-law form $N(\varepsilon) \propto 
|\varepsilon |^{p}$ with $p =$ 2.4, 2.2, 2.2, and 2.0 for the 
81 \%, 86 \%, 90 \%, and 95 \% samples respectively, at 
the (different) measurement temperatures.  As $n \rightarrow n_c$ 
from below, the gap sides steepen and the width of the Coulomb gap 
around zero bias narrows.  Although the gap "shoulders" are 
soft and therefore not precisely definable, the approximate 
full width $delta$ of the parabolic-like ($p \approx$ 2) gap 
is about 2.2 mV, 1.5 mV, 1.1 mV, and 0.6 mV and for the 81 \%, 
86 \%, 90 \% and 95 \% samples respectively.   By contrast, both 
the metallic samples show a significantly large nonzero conductance 
at $V = 0$, with a sharp dip~\cite{comment1} around zero bias that 
is well fit to a square-root form $N(\varepsilon) = 
N(0)[1+(|\varepsilon |/\delta )^{1/2}]$ at low bias.  Fits to this 
form give $\delta \approx$ 0.25 m$e$V and 0.4 m$e$V 
for the 110 \% and 105 \% samples, respectively.  Normalizing the 
metallic data to $G$(+50 mV), the ratio of zero-bias 
conductances is $G_{110}(0)/G_{105}(0) =$ 1.7, so that the 
more metallic sample has a larger DOS at the Fermi level, 
as expected.

\begin{figure}[H]
\includegraphics[width=\linewidth]{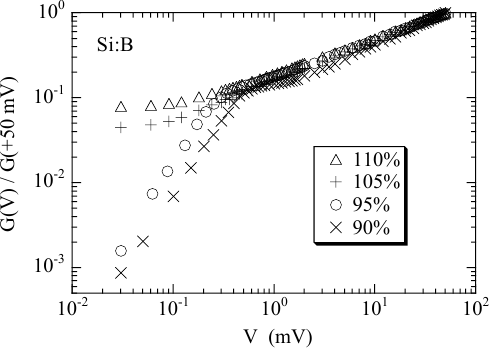}
\caption{Extended bias range tunneling conductance spectra for 
four Si:B samples, two metallic (110 \% and 105 \%) and two 
insulating (95 \% and 90 \%), all taken at temperature $T$ = 0.1 K. 
The data are  plotted on a log-log scale and are normalized to the
conductance at  +50 mV.}
\label{G(V)log}
\end{figure}

    Figure~\ref{G(V)log} compares tunneling conductance spectra 
over an extended bias range (0.03 to 50 mV) for two metallic and two 
insulating samples 5 \% and 10 \% above and below $n_{c}$, all at 
0.1 K.  This log-log-plot clearly reveals a characteristic energy 
scale of approximately 0.5 to 1 mV.  Below $\sim$ 0.5 mV, 
an obvious distinction can be made between metallic 
and insulating samples.  The low-bias data for the metallic samples 
approach $V = 0$ with a thermally rounded square-root shape and 
have a nonzero $G(V=0)$.  By contrast, the insulating samples show 
a quasi-parabolic Coulomb gap depletion of the low-energy DOS.  
The most important feature of Fig.~\ref{G(V)log} is that from $\sim$ 1 mV 
to 50 mV, the conductance spectra for both metals and insulators 
are essentially indistinguishable.  The high-bias tunneling 
conductance common for both insulating and metallic samples 
follows a functional form $G(V) \propto V^{m}$ where $m$ is between 
0.43 to 0.47, with no correlation of $m$ with dopant concentration 
in the range of samples studied.  Thus above a characteristic 
energy a DOS measurement cannot differentiate between 
metallic and insulating states.  A similar dependence was reported 
by Hertel, {\it et al.}~\cite{Hertel} in tunneling studies of 
metallic NbSi alloys near the MIT.  They found that the tunneling 
conductance showed a square-root cusp at low energies that turned 
over to a slower than square root dependence at higher energies.  
In NbSi $m$ was measured to be closer to 1/3.

    While all the junctions used SiO$_2$ barriers of roughly equal 
thicknesses, the common high-bias conductance behavior for metals 
and insulators is very unlikely to be a barrier transmission
artifact for two reasons.  First, as we already mentioned above,
the maximum  bias energy is far below the SiO$_2$ barrier height, so
that finite  voltage barrier distortions are negligible.  Second,
this behavior is  robust; it has been observed in over 25 junctions
on Si:B of varying  dopant densities, with junction resistances
($dV/dI$ at +1 mV at 1.2 K)  ranging from 60 k$\Omega$ to 850
k$\Omega$ over the same junction  area.  This order-of-magnitude
variation in junction resistance is a  consequence of uncontrolled
variations in barrier thickness and  purity.  The fact that a common
high-bias conductance is observed  despite such differences in
barrier properties indicates that the  conductance form in
Fig.~\ref{G(V)log} results from the DOS, not the barrier.  

\begin{figure}[H]
\includegraphics[width=\linewidth]{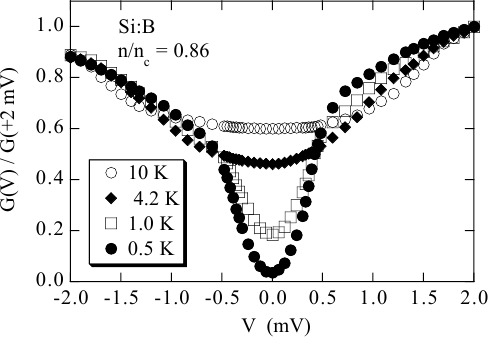}
\caption{Low-bias tunneling conductance-voltage spectra for the 
86 \% Si:B sample at several different temperatures.  The spectrum 
at each temperature is normalized to the conductance at +2 mV.  
Ordinary thermal broadening has not been deconvoluted.}
\label{G(V)temp}
\end{figure}

    Figure~\ref{G(V)temp} show details of the temperature 
dependence of the tunneling conductance for the 86 \% sample from 10
K to 0.5 K.   Ordinary thermal broadening ({\it i.e.} broadening due
solely to convolution with the $\partial f/\partial (eV)$ term in 
Eq.~(\ref {tunnelconductance}) has not been deconvoluted from 
these data.  At 10 K the tunneling conductance is essentially 
constant within $\pm$ 1 mV of $V = 0$.  As temperature decreases, 
the opening of a roughly quadratic Coulomb gap is evident.  The 
gap has its own temperature dependence (apart from the 
ordinary thermal broadening), becoming both wider and deeper as $T$ 
decreases.  The zero-bias DOS $N(V=0,T)$ goes approximately as 
$T^{2}$ at low temperature, turning over and closing ({\it i.e.} 
$G(0)/G(+2mV) \sim$ 1) near 7 K. 

\section{Scaling Ansatz for the Coulomb Gap}
\label{Scaling}
\begin{figure}[H]
\includegraphics[width=\linewidth]{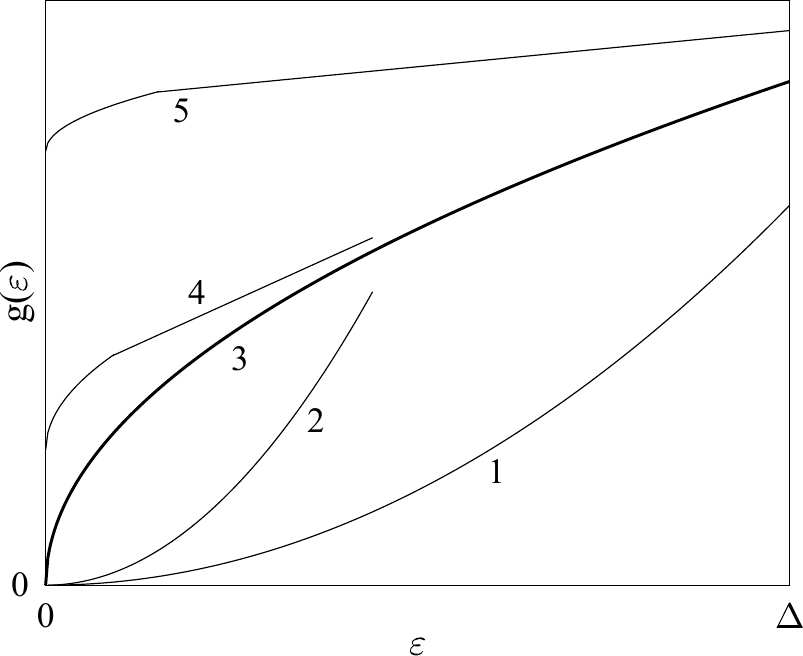}
\caption{Schematic plot of the DOS as a function of energy for 
different $n$: 1) $n \leq n_c/2$, 2) $n < n_c$ and $n_c - n \ll 
n_c$, 3) $n = n_c$, 4) $n > n_c$ and $n - n_c \ll 
n_c$, and 5) $n \geq 2n_c$.}
\label{backbone}
\end{figure}

    Coulomb interactions in a disordered insulator are known to 
cause a correlation gap in $N(\varepsilon )$.  The original ES 
derivation~\cite{ES75} for the Coulomb gap shape was given for a 
classical disordered system of point-like localized electrons 
repelling each other via the Coulomb interaction $U(r) = 
e^2/\kappa r$.  This derivation was based on a stability 
criterion for the ground state with respect to transfer of an 
electron from an occupied state $i (\varepsilon_i < 0)$ at 
position $r_i$ to an empty state $j (\varepsilon_j > 0)$ 
at position $r_j$, with:
\begin{equation}
\varepsilon_j-\varepsilon_i - U(r_{ij}) > 0,
\label{inequality}
\end{equation}
where $\varepsilon_i$ and $\varepsilon_j$ are one electron energies 
relative to the Fermi level.  If states $i$ and $j$ are within 
energy $\varepsilon$ of the Fermi level, {\it i.e.} $\varepsilon_j$, 
$|\varepsilon_i | < \varepsilon$, then their typical separation in 
space $r(\varepsilon) = e^{2}/\kappa \varepsilon$ is large if 
$\varepsilon$ is small.  The DOS $N(\varepsilon) \sim 
d(r(\varepsilon)^{-3})/d\varepsilon$ then has a soft gap around 
$\varepsilon =$ 0.  In 3-D the result is:
\begin{equation}
N(\varepsilon) = \frac{3}{\pi} \frac {\kappa^{3}\varepsilon^{2}}
{ e^{6}},
\label{DOS1}
\end{equation}
In the deep insulator, this quadratic gap extends until the 
non-interacting value $N_0$ is reached (Fig.~\ref{backbone}).  The
Coulomb gap of Eq.~(\ref {DOS1}) leads directly to the ES form for
the variable-range hopping (VRH) conductivity

\begin{equation}
\sigma = \sigma_0e^{-(T_0/T)^{1/2}},
\label{ESlaw}
\end{equation}
where
\begin{equation}
T_0=C{1\over k_B}{e^2\over \kappa\xi},
\label{To}
\end{equation}
and $C \simeq 2.8$ and $\xi$ is the localization length.

    The original ES theory applies to a disordered system far into 
the insulating side of the MIT, However, the vast majority of 
experimental observations of ES hopping use samples doped within 
10 \% to 50 \% of $n_c$.  Therefore, a description is required of 
how the Coulomb gap changes as $n_c$ is approached from below.  
One can formulate a phenomenological scaling ansatz which relates 
the DOS in the insulating state to other critical quantities 
using the ES argument.  The two quantities needed to describe both 
disorder and Coulomb interactions near the MIT are $\xi$ and 
$\kappa$.  In a doped semiconductor, when $n \rightarrow n_c$ 
from below, $\xi$ and $\kappa$ diverge with the decreasing 
parameter $(1 - n/n_c)$ as:
\begin{equation}
\xi (1 - n/n_c)  = a(1 - n/n_c)^{-\nu}
\label{xi}
\end{equation}
\begin{equation}
\kappa (1 - n/n_c)  = \kappa_0 (\xi/a)^{\eta -1 } = \kappa_0
(1 - n/n_c)^{-\zeta}
\label{kappa}
\end{equation}
where $a = n_c^{-1/3}$ is the average distance between dopants at 
$n= n_c$, $\zeta =\nu (\eta - 1)$, 
and $\nu$ and $\eta$ are scaling exponents, the same as 
used by McMillan.~\cite{MM81} (McMillanуs scaling analysis for 
metals does not calculate explicitly the values of these 
exponents but does yield the restriction 1 $< \eta <$ 3.  
Experiments~\cite{Hess82} have shown that $\zeta \sim 1$.)  
Since $\kappa$ diverges near the MIT, the Coulomb interaction 
becomes weaker and the Coulomb gap becomes steeper and narrower.  
Therefore the VRH conductivity increases due to both the increase 
of the hopping rate at a given distance following Eq.~(\ref {xi}), 
and the increase in the DOS.  This is reflected in 
Eq.~(\ref {To}), where $T_0$ tends to zero as the MIT is 
approached from below: 
\begin{equation}
k_B T_0(n/n_c) \sim  \Delta (1 - n/n_c)^{\eta \nu}
\label{To(tau)}
\end{equation}
where $\Delta = e^2/\kappa_0a$.  If $\eta \approx 2$ and $\nu 
\approx 1$, one gets $\eta \nu \approx 2$ which agrees reasonably 
with DC transport measurements in many materials.~\cite{Zab83,Zhang}  
This gives important indirect evidence that the DOS is given by 
Eq.~(\ref {DOS1}) with $\kappa$ provided by Eq.~(\ref {kappa}).  
However, DC transport is determined entirely by a relatively small 
range of energies $\varepsilon \leq k_B T_0$, so that the widest 
possible band of energies which contribute to VRH conductivity is 
$k_B T_0$.  Since $k_B T_0$ goes to zero near $n_c$, conductivity 
data contain no information about the DOS over a large range 
of energies covering 
\begin{equation}
 k_B T_0 \leq \varepsilon \leq \Delta
\label{range}
\end{equation}
In other words, in transport experiments we need only consider 
energies smaller than $k_B T_0$ or, equivalently, distances larger 
than $\xi$.  Only at such distances can the hopping probability be 
considered exponentially small and $\kappa$ be considered 
independent of length scale.

    Tunneling spectroscopy can give direct information about the 
DOS in the whole range $\varepsilon \leq \Delta$, which forces us to 
account for interactions at distances smaller than $\xi$.  
This is accomplished by introducing spatial dispersion into 
$\kappa$ at distances $r \ll \xi$ or, in terms of wavevector $q$, 
at $q \xi \gg 1$: 
\begin{equation}
\kappa \sim \kappa_0 ({r/ a} )^{(\eta -1)} \sim
 \kappa_0 ({qa} )^{(1 -\eta)}~~~~    (r \ll \xi)
\label{kappa(q)}
\end{equation}
At $r = \xi$ Eq.~(\ref {kappa(q)}) matches  Eq.~(\ref {kappa}), which 
is valid for $r \gg \xi$.  Eq.~(\ref {kappa(q)}) means that at $r \ll 
\xi$: 
\begin{equation}
U(r) = \Delta (a/r)^{\eta},~~~~   (r \ll \xi)
\label{potential}
\end{equation}
(One sees from this expression that the index $\eta$ is identical
to the standard dynamic scaling exponent $z$.)

Our goal now is to obtain the DOS throughout the energy range 
$k_B T_0 \leq \varepsilon \leq \Delta$ by repeating the ES argument 
used to derive Eq.~(\ref {DOS1}).
For $r \ll \xi$, we can use the potential in the form of 
Eq.~(\ref {potential}) only if we create wave packets of the size 
$r$ or smaller from states of the size $\xi$.  Such a packet costs 
$\hbar D(r)/r^2$ of additional "localization" energy, where $D(r)$ 
is the diffusion coefficient at length scale $r$.  We argue below 
that this energy is smaller than the energy gain per particle given 
by Eq.~(\ref {potential}) and thus allows us to proceed with the ES 
argument.  Using Eq.~(\ref {inequality}) and 
Eq.~(\ref {potential}) we calculate an average distance 
$r(\varepsilon) = a(\Delta /\varepsilon)^{1/\eta}$ 
between electron and hole states in the band of energies of width 
$\varepsilon$ around $\varepsilon_F$.  Then using Eq.~(\ref {DOS1}) 
we obtain a "critical" DOS:
\begin{equation}
N_c(\varepsilon) \simeq  a^{-3} \Delta^{-1}
({\varepsilon}/{\Delta})^{{3\over\eta}-1},
\label{DOS2}
\end{equation}
At finite $\xi$, $N_c(\varepsilon)$ matches the quadratic part 
of Eq.~(\ref {DOS1}) at $\varepsilon = k_B T_0(n)$.  Thus the width 
of the parabolic gap is: 
\begin{equation}
\delta \simeq k_B T_0(n) \sim \Delta (1 - n/n_c)^{\eta \nu}
\label{delta}
\end{equation}
Right at the transition (where $[1 - n/n_c ] = 1/\xi = 0$) 
Eq.~(\ref {DOS2}) describes the DOS at all energies $\varepsilon \
leq \Delta$ (which is why we call this DOS critical).  Thus at the very 
transition this scaling model has only one unknown index, $\eta$. 
If $\eta \approx$ 2, Eq.~(\ref {DOS2}) gives $N(\varepsilon) 
\propto \varepsilon^{1/2}$.  Critical and near-critical behavior 
of the DOS at $n = n_c$ is shown in Fig.~\ref{backbone} by a thick 
line.  This curve plays the role of a "backbone" from which 
$N(\varepsilon)$ deviates down on the insulating side of the transition ($n < n_c$) and up on the metallic side $(n > n_c$) 
near $\varepsilon = 0$ ({\it i.e.} at distances $\gg \xi$).  

Here we can connect our paper with the standard theory of quantum phase transitions~\cite{Sondhi}. This theory introduces characteristic scales of length $\xi$ and of time $\tau_\xi = h/\varepsilon_\xi$ where $\varepsilon_\xi$
is characteristic energy. At the transition both characteristic scales diverge, so that $\tau_\xi \propto \xi^z$, where z is the dynamic scaling exponent. in our case $\varepsilon_\xi =\delta =k_B T_0(n)$ and  $\tau_\xi = \propto \xi^\eta$. Thus, our exponent $\eta$ is the same as the standard dynamic scaling exponent z.

These arguments have proceeded from the insulating side.  The 
behavior of the Hartree Coulomb gap on the metallic side can also 
be estimated in the language of the dielectric constant, 
wave packets, and the ES argument.  At $q \xi \gg 1$ ($r\ll\xi$ and large energies) there is no difference between insulating and metallic phases 
so that Eq.~(\ref {kappa(q)}) is also valid for $n > n_c$.  A 
qualitative difference between metallic and insulating phases 
appears only when $q \xi \ll 1$, where instead of the scale 
independent $\kappa$ of Eq.~(\ref {kappa}) one gets metallic
scaling:~\cite{MM81,Imry82} 
\begin{equation}
\kappa(q) \sim \kappa_0 ({a/\xi})^{(1 -\eta)}(q\xi)^{-2}~~~
    (q\xi \ll 1)
\label{kappa(qq)}
\end{equation}
This divergence of $\kappa(q)$ at small $q$ leads to exponential 
screening of the interaction at distances $r \geq \xi$: 
\begin{equation}
U(r) = (e^{2}/\kappa_0 r)(a/\xi)^{\eta - 1} \exp( -r/\xi),~~~~
   (r \gg \xi)
\label{yukawa}
\end{equation}
The role of such screening was studied in Ref.~\onlinecite{ES84}.  
In the first approximation it leads to smearing of $N(\varepsilon)$ 
at the scale of $\delta = U(\xi ) = k_B T_0$.  Thus in the 
metallic state $N(\varepsilon)$ becomes finite at $\varepsilon 
= 0$: 
\begin{equation}
N(0) = A a^{-3} \Delta^{-1}
|n/n_c - 1|^{\nu (3 - \eta )},~~~~(n > n_c )
\label{DOS3}
\end{equation}

    The above expressions are all obtained in the Hartree 
approximation (with excluded self-interactions).  In this broader 
sense, all these depletions of the DOS can be called Coulomb 
gaps.  However, at very low energies in the metallic state, 
the Hartree approach fails and exchange interactions become 
dominant, leading to a square-root AA dip in the DOS near 
the Fermi energy.~\cite{AA79}  We can find the limits of the 
Hartree approximation if we compare the energy loss due to 
creation of small wave packets with the energy gain due to creation 
of a Coulomb gap.  The first quantity is $\hbar D(r)/r^2$ 
and the second is just $\varepsilon$.  Their ratio: 
\begin{equation}
Q = \hbar D(r)/r^{2} \varepsilon = \hbar \sigma (r)/r^{2} \varepsilon
 N(\varepsilon) e^2 = G_c/r^{3} \varepsilon N(\varepsilon)
\label{ratio}
\end{equation}
where $\sigma (r) = G_c e^2 /\hbar$ is the conductivity in a 
scale $r$, and $G_c$ is the critical dimensionless conductance.  
At the very transition $r(\varepsilon) = a(\Delta /\varepsilon)^{1/\eta}$, 
so that Eq.~(\ref {ratio}) with the help of Eq.~(\ref {DOS1}) 
shows that $Q \leq 1$, meaning the Hartree approximation is 
still reasonable.  This justifies the backbone DOS 
Eq.~(\ref {DOS2}) at energies $\varepsilon > \delta$ on both 
sides of the transition.  

    On the other hand, at $\varepsilon_c < \delta$ in the 
metallic state the Hartree approximation fails, wave packets 
overlap, and an AA-like theory based on exchange interactions 
becomes valid.  As a result, a negative square root of comparable 
amplitude should be added to Eq.~(\ref {DOS3}).  

    Now we would like to compare our results with those obtained 
by McMillan~\cite{MM81}.  In his work the Hartree interaction was 
neglected and the metallic side of the transition was strongly 
emphasized.  He did not present an explicit form of low energy 
behavior of DOS on the insulating side and did not mention 
that it is described by notion of the Coulomb gap.  Nevertheless 
his estimate of the "correlation gap" width $\Delta$ is in 
agreement with our width $\delta = k_B T_0$ of the parabolic Coulomb 
gap.  In the metallic phase and in the critical regime 
our results are identical to those of McMillan~\cite{MM81}.  
(Note that we use same $\eta$ and $\nu$ as McMillan, but his 
$\Delta $ is equivalent to our $k_B T_0$ or $\delta$.)

   Refs.~\onlinecite{MM81,Gefen} were criticized~\cite{Lee82} 
for using the single-particle DOS $N(\varepsilon)$ in the dielectric 
constant: 
\begin{equation}
\kappa = \kappa_0 (1 + {{4\pi N(\varepsilon) e^{2}} \over{q^2}})
\label{kappa(qqq)}
\end{equation}
The standard expression for $\kappa$ of a homogeneous system uses 
the thermodynamic DOS $dn/d\mu$ in place of $N(\varepsilon)$ in 
Eq.~(\ref {kappa(qqq)}), where $\mu$ is the chemical potential.  
In an electron system with direct Coulomb interactions, $dn/d\mu$ 
is believed to have no gap near the Fermi level, in 
contrast to $N(\varepsilon)$.  So which of the two DOS enters 
Eq.~(\ref {kappa(qqq)}) is the crucial question.  This question 
is directly relevant here, since all the results in this paper appear 
to agree with the use of Eq.~(\ref {kappa(qqq)}).  Indeed, 
Eqs.~(\ref {kappa}),~(\ref {kappa(q)}), and~(\ref {kappa(qq)}) 
are self-consistently related by Eq.~(\ref {kappa(qqq)}) to 
Eqs.~(\ref {DOS1}),~(\ref {DOS2}), and ~(\ref {DOS3}) respectively.  

    We can repeat the problem in terms of the screening radius 
$r_s$ by rewriting Eq.~(\ref {kappa(qqq)}) as $\kappa = 
\kappa_0 + {1 / ({r_s}^{2}{q^2})}$.  If  ${r_s}^{2} = 
\kappa_0/4\pi N(\varepsilon) e^{2}$ then at small 
$\varepsilon$, $N(\varepsilon)$ is small, the radius $r_s$ is 
large, the screening is weak so the interaction is strong, and 
finally $N(\varepsilon)$ has a Coulomb gap.  For example, 
deep on the insulating side, using Eq.~(\ref {DOS1}) we get 
that at energy $\varepsilon$ the radius $r_s = e^2 /\kappa_0 
\varepsilon$, exactly equal to the average distance between 
states within $\varepsilon$ of the Fermi level.  This implies 
that screening self-consistently does not destroy the Coulomb gap.  
The other option argued for in Ref.~\onlinecite{Lee82} starts 
from the DOS $dn/d\mu$.  If $dn/d\mu$ is large, ${r_s}^{2} 
= \kappa_0/(4\pi e^{2} dn/d\mu)$ is small and we get a short range 
interaction which does not lead to the Coulomb gap.   This result is 
inconsistent with the ES argument, computer simulations, and numerous 
observations of the ES law for VRH conductivity.    
Why, then, does the conventional definition of the screening 
radius fail?  

    An answer to this question for the deeply insulating phase 
was suggested in Ref.~\onlinecite{Baranovskii}.  While the correct 
expression for $\kappa$ contains $dn/d\mu$, in a strongly 
inhomogeneous system with localized states one needs a 
{\em local\/} $\kappa$ and $dn/d\mu$ to describe the interaction 
of two particular states at the distance $r$.  The local 
$(dn/d\mu)^{-1}$ fluctuates strongly in absolute value and has 
random sign.  The inverse amplitude of these fluctuations equals 
$N(\varepsilon)$.  Thus, the local screening is determined by a 
small random sign DOS which in absolute value is of order 
$N(\varepsilon)$.  The random sign of the interaction does not 
change any estimates based on the ES argument, and this is why the 
ES argument works.  On the other hand, $(dn/d\mu)^{-1}$ 
averaged over realizations of a random potential is very small, 
so that the thermodynamic DOS $dn/d\mu$ is large and energy 
independent.  The absence of self averaging in strongly disordered 
systems is the main reason that the {\em average\/} $dn/d\mu$ 
does not determine the strength of local interactions.  
Ref.~\onlinecite{Baranovskii} discussed a classical case of a deep 
insulator.  We assume that in the critical range of the MIT 
quantum effects can be taken into account by renormalization of 
$\xi$ and $\kappa$ and by introduction of wave packets.  After that 
we have a problem similar to the classical one, but with 
renormalized interactions.  This interaction for a given 
realization of impurities is not screened.  To find $N(\varepsilon)$ 
one needs the local interaction, and the average $dn/d\mu$ is 
irrelevant.

    Let us remind the computer simulations lead to these
conclusions~\cite{Baranovskii}.  In that work the authors considered 
randomly distributed pointlike donors and acceptors, all acceptors 
being negatively charged and some donors being occupied by electrons 
(and hence neutral) and others empty (and positive).  The ground 
state  of such system is a classical realization of a Coulomb 
gap~\cite{ES75,ES84}.  Screening of a probe positive charge located at the ceter of sphere at $T = 0$ was studied numerically.  It was shown that for a given realization of impurities, the screened potential of the probe
charge is almost random in sign and its absolute value decays as
$1/r$. To understand the origin of the random sign potential for a 
single realization of potential, recall that in the ground state 
electrons are correlated in space or, in other words, positive
donors alternate with  neutral ones.  The probe positive charge
attracts a new electron to  the nearest empty donor with probability of order of 1/2.  The positive donor left behind attracts a new electron to another its
neighbor, with probability of order of 1/2 and so on.  As a result a chain of new  alternating positive and negative charges can appear.  Such a chain has  random direction and length and no symmetry.  It is clear that the  sign of
the resulting potential, for example at the end of the  chain,  is random. Huge cancellations happen  when we average this potential over many different
realizations of impurities or over different  configurations of
chains.  As a result spherical symmetry is restored and  an
exponentially small positive potential survives. This picture has 
been generalized for a description of the Rudderman-Kittel exchange 
interaction in a disordered metal.~\cite{Spivak}.

    An alternative explanation was recently suggested by Si and 
Varma,~\cite{Si} who argued that, contrary to previous 
assertions, direct Coulomb interactions do lead to a gap in the 
average $dn/d\mu$ when the density is low enough in a 2D system.  
They suggest that the incorporation of this result into the charge 
diffusion models of Refs.~\onlinecite{Fink83,Castel87,KB90} could 
generate a realistic description of the transition from diffusive 
metal to Coulomb glass insulator.

\begin{figure}[H]
\includegraphics[width=\linewidth]{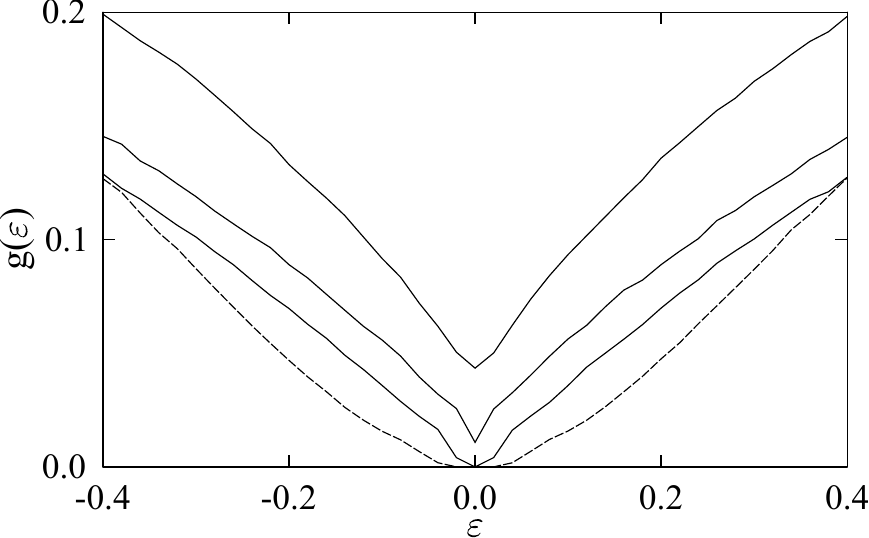}
\caption{DOS as a function of energy as obtained by numerical 
simulation for the Coulomb potential (dashed line) and for 
three model potentials corresponding to the critical range of 
the MIT (full lines) from bottom to top: $n < n_c (\xi /R = 2)$, 
$n = n_c$, and $n > n_c (\xi /R = 2)$.}
\label{numerical}
\end{figure}

    Finally, Fig.~\ref{numerical} presents numerical calculations 
of DOS spectra of the impurity band model obtained by computer
minimizations of the  total energy of the system in the framework of
our scaling ansatz  model.  To describe the MIT at $n < n_c$, $n =
n_c$, and $n > n_c$,  we use Coulomb potentials modified by 
Eqs.~(\ref {kappa}),~(\ref {kappa(q)}), and ~(\ref {kappa(qq)}) 
respectively.  For concreteness, we assume 
$\eta = 2$ and $U(r) = (R/\rho)^2$ for $n = n_c$; $U(r) = 
(\rho^2 +\xi^2)^{1/2} R/\rho^2$ for $n < n_c$; $U(r) = R^2/[\rho 
\xi (\exp \rho/\xi - 1)]$ for $n > n_c$  Here $R = n_c^{-1/3}$ is 
the average distance between donors at the transition, $U$ is 
measured in units of $e^2/{\kappa_0 R}$, and $\rho^2 = r^2 +
(R/2)^2$.  The $(R/2)^2$ term in the last expression is significant
only at small $r$ and is introduced as a large energy cut-off (or
alternatively as  a small wavelength cut-off at distances $a \sim
R/2$ at $n_c$).   The algorithm of the simulation is described in
Ref.~\onlinecite{ES84}.  The energy reference is adjusted  for each
realization of impurity coordinates so that the chemical  potential
is zero, and the DOS is averaged over 10000 realizations.   In
Fig.~\ref{numerical}, the insulating and metallic DOS are computed 
for states equidistant from the transition, using $\xi = 2R$.   We
see that results of calculation are in a reasonable  agreement with
the schematic scaling plot of Fig.~\ref{backbone} and  the
experimental data of the previous section.

\section{Analysis of the Tunneling Data}
\label{analysis}
   There are clearly remarkable similarities between the 
experimental data of Fig.~\ref{G(V)log} and the scaling ansatz 
results in Fig.~\ref{backbone}.  In fact, most major features of the
tunneling  measurements can be accounted for by the simple scaling 
analysis of the preceding section.  It is obvious from 
Fig.~\ref{G(V)log} that the high-bias conductance common to both
localized  insulator and disordered metal plays the role of the high
energy  backbone DOS obtained from the scaling ansatz by combining 
the dielectric constant of Eq.~(\ref {kappa(q)}) with the 
Coulomb gap of Eq.~(\ref {DOS1}).  The distinction between 
metal and insulator becomes clear only when the DOS departs from 
the backbone at low energies.  From the data, this departure 
occurs near an energy scale of about 0.5 m$e$V.  

    While the scaling ansatz cannot calculate the value of the 
exponent describing the power law backbone DOS, the experimental 
data does yield this number and the value of the associated scaling 
exponent.  According to Eq.~(\ref {potential}), the high-energy 
backbone DOS near the critical density for both metals and 
insulators depends on a single exponent $\eta$ and should go 
as $N_c (\varepsilon) \propto \varepsilon^{\frac{3}{\eta} - 1}$.  
From the data of Fig.~\ref{G(V)log}, in the bias range 1 mV $\leq V 
\leq$ 50 mV we obtain $G(V) \propto V^m$ with values for 
the exponent $m = 0.45 \pm 0.2$ covering all the samples measured.  
This gives $\eta = 3/(1 + m) =  2.1$, which is certainly within 
the theoretically required  bounds $1 < \eta < 3$ and in fact is 
essentially the same as the  value $\eta = 2$ reported by Hertel, 
{\it etal.}~\cite{Hertel}  in barely metallic NbSi alloys.  Since 
this value for $\eta$ was  obtained from the high-energy DOS, 
where metals and insulators  share a common DOS character, and from 
many different samples  both below and above $n_c$, we believe it is 
a reliable value.   We do wish to emphasize that, in our 
interpretation, the approximately  square-root behavior of the 
DOS at these high energies is  unrelated to the AA exchange 
correction, which is  important only at energies $\varepsilon < 
\delta \approx$ 0.4 m$e$V in the metallic state.

\begin{figure}[H]
\includegraphics[width=\linewidth]{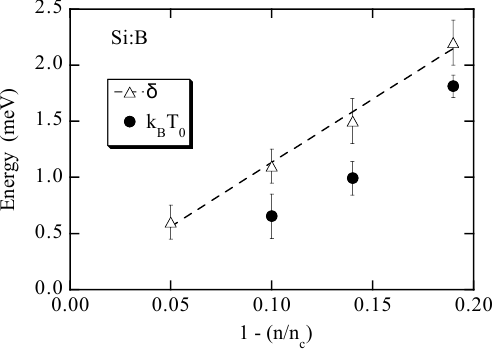}
\caption{Full width $\delta$ of the parabolic Coulomb gap in four 
insulating samples plotted against $(1 - n/n_c)$.  The dashed line 
is a linear fit.}
\label{delta(n)}
\end{figure}

    The exponent $\nu$ can also be examined, though less definitively 
than $\eta$.  On the metallic side of the MIT this can be done using 
Eq.~(\ref {DOS3}).  The measured ratio of zero-bias conductances 
for the two metallic samples $G_{110}(0)/G_{105}(0) = 1.7$. 
Relating this  conductance ratio to the ratio of the DOS and using
Eq.~(\ref {DOS3}),  we obtain $\nu (3 - \eta) \approx$ 0.77.  Taking
$\eta \approx 2$ from  the preceding analysis, this gives $\nu
\approx$ 0.77.   Alternatively, $\nu$ can be estimated independently
using data  from the insulating samples.  Fig.~\ref{delta(n)} shows
the measured Coulomb gap  widths for four insulating samples plotted
against normalized dopant density.   The data fit a linear function
very well with nearly zero intercept,  {\it i.e.} $\delta (n)
\propto (1 - n/n_c)^{\gamma}$ where $\gamma \approx  1.0$.  From
Eq.~(\ref {delta}), we have $\gamma = \eta \nu$, so that if 
$\eta =  2$, then the gap width data imply that $\nu = 0.5$.  
Fig.~\ref{delta(n)} also shows that, in agreement with Eq.~(\ref
{delta}), the  values of $T_0$ measured from  DC conductivity in the
more insulating  samples, where the ES hopping law is followed most
rigorously, are  of the same order as the  corresponding values of
$\delta$.

    These values for $\nu$ obtained from the tunneling data can be 
compared to the independent data for $\kappa (n)$ shown 
in Fig.~\ref{kappa(n)}.  A simple power-law fit to the data in that 
figure gives $\zeta \approx 0.71 = \nu (\eta - 1)$.  Again, if
$\eta$ is taken to be 2, then 
$\nu \approx 0.71$.  This value is comparable to the values obtained 
from the DOS data analysis.  Therefore, we conclude that the DOS
data support  values for the scaling exponents of $\eta \approx 2$
and $\nu$ somewhere  between 0.5 and 0.8, at least within the range
of $n/n_c$ covered by the  samples used.

    The exponent $\nu$ is commonly measured by examining the very 
low temperature DC conductivity $\sigma (T \rightarrow 0)$ as a 
function of doping and fitting the data to the form $\sigma(0) 
\sim (n - n_c)^\nu$ for $n$ slightly above $n_c$.  For nominally 
uncompensated doped silicon (Si:B,~\cite{Dai91} 
Si:P,~\cite{Rosenbaum81} and Si:As~\cite{Shafarman89}), many such
transport experiments have  reported values for $\nu$ between 0.5
and 1, generally closer to 0.5  than to the $\nu \approx 1$ observed
in most other disordered  conductors and expected theoretically.  In
Si:B, for example, the  exhaustive very low temperature conductivity
measurements reported  in Ref.~\onlinecite{Dai91} indicate that $\nu
= 0.65$.  Thus our analysis  of the tunneling data yield a value for
$\nu$ at least consistent  with that obtained from transport
measurements.  However, the  interpretation of the transport data
leading to $\nu \approx 0.5$  in
Refs.~\onlinecite{Dai91,Rosenbaum81,Shafarman89} has recently been 
questioned.~\cite{Stupp,Shlimak}  There is some evidence that if 
only samples within $\sim$ 1 \% of $n_c$ are used, then transport 
measurements yield $\nu \approx 1$ instead.  In fact, stress  tuning
of a Si:B sample~\cite{Bogdanovich1} across the MIT has  been
reported to give $\nu$ as high as 1.6, although  earlier stress
experiments~\cite{Paalanen} on Si:P reported 
$\nu$ = 0.5.  

    An estimate for the value of $\xi$ can also be extracted from 
the data.  Using Eq.~(\ref {To}) combined with the measured values 
of $\kappa$ and $T_0$ in Figs.~\ref{kappa(n)} and~\ref{delta(n)} 
respectively, we obtain $\xi \approx$ 25 nm, 30 nm, and 40 nm for 
the 81 \%, 86 \%, and 90 \%  samples, respectively. These values 
are in reasonable agreement with Eq.~(\ref {xi}) using 
$\nu \sim 0.7$ and length $a = 7.5$ nm, which is close to the 
average distance between impurities at $n_c$, 6.3 nm. 

    It is necessary to discuss whether the scaling ansatz of 
Section~\ref{Scaling} should be really applicable to the experimental 
situation of Section~\ref{Tunneling}. First there is a  problem 
with theweak compensation  of the Si:B samples used.  Indeed  the
theoretical discussion starts  from the picture of a classical
Coulomb gap.   In the impurity band  of a lightly doped
semiconductor, the Coulomb gap is most pronounced  for degrees of
compensation (ratio of minority  to majority impurity 
concentration) comparable to 1/2.  (In this case  a large number of
impurities are charged, but the peak of the DOS is still not too
broad.~\cite{ES84})  Then with increasing $n$ the Coulomb gap
evolves  according to the  predictions of Sec. 3.  In uncompensated
lightly  doped semiconductors, all majority impurities are occupied
in the  ground state and therefore are neutral.  There is a large
Mott-Hubbard  gap which separates occupied states from empty states
of a second  electron on an impurity.  Disorder is very weak because
there are  no random charges.  So the Mott-Hubbard gap at the Fermi
level of a  lightly doped uncompensated semiconductor is a real
hard  gap as  opposed to the soft Coulomb gap.  Why then can we then
talk about a  Coulomb gap near the MIT?  Bhatt and Rice~\cite{Bhatt}
pointed out  that at large impurity concentrations $n \sim n_c$ a
new phenomenon  which they called self-compensation can take place. 
Because of large  overlap and strong positional disorder of their
wave functions,  some clusters of impurities can have large affinity
to electrons and  can ionize other  clusters. As a result charges
will appear and a  random potential will close  the Mott-Hubbard gap
which is already  narrowed at these concentrations.   Then our
theory, developed for  a compensated semiconductor can work even for
a nominally uncompensated  one.  Similar tunneling experiments on
compensated samples would be  very important to find out  whether
they provide a similar DOS and  whether it agrees with our theory.

    A second problem is related to the role of the Al electrode 
in the screening of the long-range Coulomb interactions.  One can 
think that tunneling electrons typically penetrate a distance 
$\xi$ into a semiconductor. The characteristic length 
$r(\varepsilon)$ of interactions which  determines the DOS at
$\varepsilon \gg \delta$  (square root region) is smaller than
$\xi$. This means that the  large square root region  is robust with
respect to screening by a  metal gate.  However, the parabolic gap
results from the Hartree  interaction between charges  separated by
distances $r >\xi$.   Therefore, the Al electrode could produce
substantial image charge  screening of this interaction.  Such 
screening was studied numerically~\cite{Cuevas} for the case of a
classical Coulomb  glass and a metal electrode placed directly on
its surface.   These computations show the bulk parabolic Coulomb
gap near the  surface can be closer to a linear one, rather than
quadratic one.  In  our experiment, electrode screening may be
reduced by two factors.   First, the Al electrode is backed off the
semiconductor surface  by roughly 1.5 to 3 nm of SiO$_2$, as
described in Section~\ref{Tunneling}.   Second, the Schottky 
barrier between the Al and Si can be as  wide as 20 nm.  The
cumulative effect of these two barriers is to  increase the distance
between the  charges and image charges to  distances $\sim \xi$.  
However, even if the gate is backed off,  the parabolic gap should
still be  expected to fail at small  enough energies.  Surprisingly,
the parabolic gap is more robust  than expected.  There is strong
agreement between the power law  for the Coulomb gap as measured by
tunneling and by bulk DC  transport, which is not affected by
electrode screening because  the sample is macroscopically thick
($\sim$ 250 to 300
$\mu$m) so that most of current flows far from the Al contacts.  
If the Coulomb gap is  described by $N(\varepsilon) \sim 
\varepsilon^p$, then the ES model gives a  hopping exponent $s =
(p+1)/(p+4)$ for  the conductivity $\sigma = \sigma_0
\exp\{-(T_0/T)^s\}$.   For the  86 \% sample at 0.5 K, a power-law
fit to the tunneling data gives 
$p = 2.2$, so the predicted ES value for $s =  0.52$.  Transport 
measurements on the same sample in the same temperature range give 
a value $s = 0.51 \pm 0.02$.  The close consistency o the tunneling 
and transport results indicates that, to a reasonable accuracy, 
the tunneling DOS reflects the bulk characteristics.  

    Finally, we should mention that the definition of DOS discussed 
in section 2 does not strictly coincide with the definition of the 
tunneling DOS measured experimentally.  Section~\ref{Tunneling} 
dealt with the bulk DOS of charged  excitations (electronic
polarons). For a deep  insulator these excitations were  introduced
in  Ref.~\onlinecite{ES84}.  Electronic polarons are responsible for
ES  VRH.  On the other hand  tunneling experiments measure the 
one-electron DOS.  For a deep  insulator, a substantial difference
between these two DOS in  the limit of small energies was originally 
predicted in Ref.~\onlinecite{ES84}, but  was not observed in
numerical simulations. This issue was clarified in Ref.~\cite{Efros2011}.
\section{Summary}
\label{summary}
   Electron tunneling measurements of the single-particle DOS have 
beenmade on Si:B crystals ranging from 81 \% to 110 \% of the
critical density of the MIT.  At low energies ($\varepsilon \leq$
0.5 m$e$V), non-metallic samples show an approximately quadratic
soft Coulomb gap across the Fermi level, while metallic samples show
a square-root cusp.  At higher energies, (up to 50 m$e$V), both
insulating and metallic samples show a common DOS behavior,
$N(\varepsilon) \sim \varepsilon^m$. with $m$ slightly less than
1/2.  These features of the data can be understood within the
framework of a scaling ansatz of the approach to the MIT from the
insulating side.  By extending the semi-classical Efros-Shklovskii
derivation of the Hartree Coulomb gap to include a spatial
dispersion of the dielectric constant at small length scales, it was
shown that at short length scales ($r \ll \xi$), or high energies,
spatial dispersion of $\kappa$ gives rise to a common DOS that
increases as a power law in both metals and insulators.  This is
distinct from the Altshuler-Aronov low-energy square-root cusp in
the DOS arising from exchange correlations in weakly disordered
metals.  Only at long length scales, or, equivalently, low energies,
does a quadratic Coulomb gap become apparent in insulators, while
metals take on a non-zero DOS across the Fermi level.  Finally, the
tunneling results and the scaling ansatz suggest that a
semi-classical approach may be used to treat Coulomb interaction
effects on the approach to the MIT from the insulating state. Our scaling ansatz is supported by two recent publications~\cite{Markus,Gornyi}. 

%
\acknowledgements

We thank J. K. Wescott for help with the SiO$_2$ growth, and A. A.
Koulakov, M. M. Fogler and A. L. Efros for useful discussions of
theoretical questions.  Work  at U.Va. was supported by NSF grant
DMR-9700482 and by a Cottrell Award  from the Research Corporation.  Work
at U.Mn. was supported by NSF grant  DMR-9616880
%


\begin{references}
\bibitem{comment1} The non-analytic point at $V = 0$ is always 
thermally broadened.
\end{references}
\end{document}